\documentclass[10pt,twocolumn]{article}
\usepackage[margin=0.75in]{geometry}
\usepackage[caption=false,font=footnotesize]{subfig}
\usepackage{graphicx}
\usepackage{cite}
\usepackage{array}
\usepackage[cmex10]{amsmath}
\usepackage{amssymb}
\usepackage[capitalise]{cleveref}
\usepackage{booktabs}
\usepackage{enumitem}
\usepackage{acronym}
\usepackage{csquotes}
\usepackage{balance}
\usepackage{color}

\newcolumntype{C}[1]{>{\centering\let\newline\\\arraybackslash\hspace{0pt}}m{#1}}

\newacro{FA}{full-adder}
\newacro{LUT}{lookup table}
\newacro{GP}{General Purpose}
\newacro{SoC}{System on Chip}
\newacro{CNOT}{Controlled NOT}
\newacro{GSS}{Ground State Spin}
\newacro{RCA}{ripple-carry adder}
\newacro{FSM}{Finite State Machine}
\newacro{ALU}{arithmetic logic unit}
\newacro{RTL}{Register-Transfer Level}
\newacro{LSI}{Large Scale Integration}
\newacro{LIF}{Leaky Integrate and Fire}
\newacro{MTJ}{Magnetic Tunnel Junction}
\newacro{ANN}{Artificial Neural Network}
\newacro{SGD}{Stochastic Gradient Descent}
\newacro{EDA}{Electronic Design Automation}
\newacro{FPGA}{Field-Programmable Gate Array}
\newacro{B2S}{Binary-to-Stochastic Converter}
\newacro{PRNG}{pseudo-random number generator}
\newacro{LFSR}{Linear-Feedback Shift Register}
\newacro{CDF}{cumulative distribution function}
\newacro{VDEC}{VLSI Design and Education Centre}
\newacro{PGS-D}{Postgraduate Scholarship - Doctoral}
\newacro{ASIC}{Application Specific Integrated Circuit}
\newacro{CMOS}{Complimentary Metal-Oxide-Semiconductor}
\newacro{JSPS}{Japan Society for the Promotion of Science}
\newacro{TSMC}{Taiwan Semiconductor Manufacturing Company}
\newacro{NS1e}{Neurosynaptic System, 1 million neuron evaluation platform}
\newacro{NSERC}{Natural Sciences and Engineering Research Council of Canada}
\newacro{MEXT}{Ministry of Education, Culture, Sports, Science and Technology}
\newacro{JST}{Japan Science and Technology Agency}
\newacro{PRESTO}{Precursory Research for Embryonic Science and Technology.}

\author{
Sean C. Smithson$^{1}$, Naoya Onizawa$^{2}$, Brett H. Meyer$^{1}$, Warren J. Gross$^{1}$, and Takahiro Hanyu$^{2}$\\
\normalsize $^{1}$Department of Electrical and Computer Engineering, McGill University, Montreal, QC, Canada H3A 0E9\\
\normalsize $^{2}$Research Institute of Electrical Communication, Tohoku University, Sendai, Miyagi, Japan 980-8577\\
\normalsize \texttt{sean.smithson@mail.mcgill.ca, brett.meyer@mcgill.ca, warren.gross@mcgill.ca}\\
\normalsize \texttt{naoya.onizawa.a7@tohoku.ac.jp, hanyu@riec.tohoku.ac.jp}
}
\date{}

\title{Efficient CMOS Invertible Logic \\ Using Stochastic Computing}

\begin{document}

\flushbottom
\maketitle

\begin{abstract}
Invertible logic can operate in one of two modes: 1) a forward mode, in which inputs are presented and a single, correct output is produced, and 2) a reverse mode, in which the output is fixed and the inputs take on values consistent with the output. It is possible to create invertible logic using various Boltzmann machine configurations. Such systems have been shown to solve certain challenging problems quickly, such as factorization and combinatorial optimization. In this paper, we show that invertible logic can be implemented using simple spiking neural networks based on stochastic computing. We present a design methodology for invertible stochastic gates, which can be implemented using a small amount of \acs{CMOS} hardware. We demonstrate that our design can not only correctly implement basic gates with invertible capability, but can also be extended to construct invertible stochastic adder and multiplier circuits. Experimental results are presented which demonstrate correct operation of synthesizable invertible circuitry performing both multiplication and factorization, along with fabricated \acs{ASIC} measurement results for an invertible multiplier circuit.
\end{abstract}

\section{Introduction}
\label{sec:introduction}

As the foundation for nearly all computing circuitry, digital logic is inherently unidirectional. For each given input there is one, and only one, corresponding output value (or values). However, the reverse does not always hold true; for any given output value, there may exist a number of valid input combinations that satisfy the constraints imposed by the Boolean function implemented. While a NOT gate is inherently invertible, consider the elementary AND gate: when the output is low, there exist three possible input combinations with no way of determining which is the desired. The ability to operate logic circuits in reverse would allow for a variety of desirable characteristics: 1) entire combinational circuits could be used completely in reverse (e.g. a multiplier could be used as a factorizer), and 2) partial combinations of inputs and outputs could be fed into a circuit and the remaining terminals used as outputs (e.g. an adder with the sum and a single operand as inputs, would output the subtraction thereof) \cite{Camsari:pbits_X}. In turn, hardware costs could be reduced through re-using invertible logic circuits for multiple purposes; for example an invertible adder circuit can also perform subtraction and an invertible multiplier circuit can also perform division, in addition to factorization. One example of using invertible arithmetic circuits to reduce hardware costs, or improve overall performance, could be when designing for diverse workloads; a designer may choose to implement invertible multipliers, in place of a set number of multiplier and divider circuits, in order to achieve greater hardware utilization on different workloads. Furthermore, invertible logic circuits are also of interest in the fields where reversible computing in general has demonstrated promise (cryptography, digital signal processing, and computer graphics) \cite{Saeedi:ReversibleComputing}. 

\begin{figure}[t]
	\centering
	\includegraphics[width=0.95\linewidth]{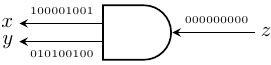}
    \caption{Inverted stochastic AND gate.}
	\label{fig:SC_AND}
\end{figure}

The basis for stochastic invertible logic is to represent inputs and outputs not as fixed values, but as probabilities of a signal being either high or low \cite{Camsari:pbits_X}. Through adopting such a strategy, when there exists many input value combinations satisfying a given output value, the output of the stochastic circuit is a random variable whose distribution has probability mass concentrated at valid values. Illustrated in \cref{fig:SC_AND} for the aforementioned case of an invertible AND gate operating in reverse; if the output is kept at $z=0$, then the inputs $\left( x, y \right)$ alternate equally ($33\%$ of the time, each) between the valid values of $\left( 0, 0 \right)$, $\left( 0, 1 \right)$, and $\left( 1, 0 \right)$.
\begin{table*}[t]
    \caption{Comparison of Logic Family Characteristics}
    \label{table:logic_compare}
    \centering
    \begin{tabular}[width=\textwidth]{ccccccc}
        \toprule
        Logic family & Forwards operation & Backwards operation & Existing \acs{EDA} tools & Deterministic & Probabilistic & All valid values \\
        \midrule
        Conventional & \checkmark & X & \checkmark & \checkmark & X & X \\
        \midrule
        Reversible & \checkmark & \checkmark & X & \checkmark & X & X \\
        \midrule
        Invertible (exotic devices) & \checkmark & \checkmark & X & X & \checkmark & \checkmark \\
        \midrule
        Invertible (\acs{CMOS})  & \checkmark & \checkmark & \checkmark & X & \checkmark & \checkmark \\
        (this work, \cite{Pervaiz:pbits_FPGA}) & & & & & & \\
        \bottomrule
    \end{tabular}
\end{table*}
We present invertible multiplier circuits which can also operate as factorizers. However, the motivation behind this work comes not only from existing applications, but also for emerging areas where we envision invertible logic circuits can make significant contributions. One logical progression of our work is extending multiplication of scalars to complete invertible matrix multipliers to speed up machine learning. Many machine learning algorithms require time- and memory-intensive algorithms to find values for parameter matrices that minimize a cost function. An invertible matrix multiplier could enable more efficient learning by simply inverting the cost function directly.

In this work we present a family of invertible logic circuits based on stochastic computing circuits as the underlying processing elements which can be easily manufactured in standard \ac{CMOS} processes. We describe the overall hardware architecture, along with the methodology we followed to design invertible multiplier (and factorizer) circuits. We also present \ac{RTL} simulation results of fully-synthesizable multiplier designs, alongside measurement test results of a fabricated invertible logic circuit. Our results demonstrate that not only can stochastic computing be used as a theoretical basis for invertible logic but also that it can be manufactured with existing methods, taking advantage of the current state-of-the-art technologies. In addition to the presented method being fully synthesizable, the nature of our work also renders it well suited for \ac{FPGA} based applications. In particular, even for applications where the underlying Boltzmann machines are implemented with exotic devices (be they analogue, or even stochastic in nature), our approach allows for prototyping of the underlying invertible logic structure on reprogrammable \ac{FPGA} fabric, which in turn may reduce overall engineering time and development costs.
Our contributions, when compared with the conventional work  (especially \cite{Camsari:pbits_X}) summarized in Table \ref{table:logic_compare}, include: (a) a circuit design that can be manufactured using standard digital CMOS with existing \ac{EDA} tools (such as Cadence and Synopsys tools), (b) the first demonstration of the fabricated invertible logic-circuit chip (hardware), and (c) smaller number of nodes (smaller sizes of Hamiltonian).
In comparison with \cite{Pervaiz:pbits_FPGA}, this work exploits stochastic computing to implement invertible logic in CMOS while traditional binary logic is used with larger Hamiltonians in \cite{Pervaiz:pbits_FPGA}.
The proposed invertible multiplier is fabricated using \ac{TSMC} 65nm \ac{GP} process in this paper, but can be of benefit in more advanced CMOS technologies.

The rest of the paper is as follows. 
Section \ref{sec:related_works} reviews related works.
Section \ref{sec:stochastic} describes the basis of stochastic computing.
Section \ref{sec:architecture} introduces the proposed invertible circuit design.
Section \ref{sec:inv_logic_circuits} presents the design methodology of Hamiltonian for invertible logic circuits.
Section \ref{sec:experimental_results} evaluates the proposed invertible logic circuits in simulation and measurement results and compares with conventional works.
Section \ref{sec:conclusions} concludes the paper.

\section{Related Works}
\label{sec:related_works}

\subsection{Invertible Logic}
\label{sec:related_inv_logic}

Previous works have explored invertible logic circuit families, however such topologies are built around the use of exotic devices, and remain non-manufacturable using standard \ac{CMOS} technologies. In the case of \ac{GSS} logic, quantum devices are required \cite{GSS1,GSS2}. 
\textit{Probabilistic spin logic} necessitates the use of naturally probabilistic switching devices (such as \ac{MTJ}-based devices) \cite{pbits_device,Camsari:pbits_X,superparamagnetic}. 
While such approaches may ultimately lead to viable technologies, a \ac{CMOS}-based alternative would not only be deployable today, taking advantage of state-of-the-art manufacturing processes, but also be more easily combined with conventional digital logic circuitry. Given that modern \ac{SoC} designs can include various types of circuitry, the ability to selectively implement those desired to also be invertible, without affecting the rest of the system, allows for a flexibility lacking in these other theoretical technologies.

\subsection{Reversible Logic}
\label{sec:related_rev_logic}

Differing from invertible logic is reversible logic, where circuits are constructed of special gates (such as \ac{CNOT} or Toffoli gates) having a direct one-to-one mapping of inputs to outputs \cite{Saeedi:ReversibleComputing}. In such circuits there are no two inputs which give the same output values. While reversible logic gates allow for circuits to be built which are invertible, they must be designed differently and do not include standard gates (such as AND or OR gates) and require different design flows \cite{Zulehner:ReversibleDesign}. While similar, that is both reversible and invertible logic circuits reconstruct inputs from a given output value, they differ at fundamental levels with differing design goals. Outlined in \cref{table:logic_compare} are the key characteristics of invertible logic when compared to other logic families. Of note, is that the invertible logic approach taken in this work can produce a drop-in replacement in existing EDA tools, and allow for designers to not only obtain a single corresponding valid input for a given output, and because of the stochastic nature of this approach, allow for the recovery of all possible valid inputs as well.
%

\subsection{Spiking Neural Networks}
\label{sec:related_spiking}

Differing from more traditional \ac{ANN} models, digital spiking neurons were developed as low power processing elements taking inspiration from biology \cite{Merolla:TrueNorth_Science}. Forgoing real-valued inputs and outputs, spiking neurons communicate through series of pulses; specifically for digital spiking neurons the pulse train communication is further simplified by replacing the analogue waveforms of biology with single-bit streams which are high (\enquote{1}) when the neuron fires, and low (\enquote{0}) otherwise. Digital spiking neurons are not only interesting due to their biological plausibility, but also because of their inherent simplicity, and potential for very low area and energy implementations. Computing the weighted sum of neuron inputs is simplified by eliminating the need for costly binary multipliers; instead, a series of addition operations are sparsely distributed over time (occurring irregularly and infrequently) \cite{Akopyan:TrueNorth}. This reduction in processing element complexity, when also combined with an event driven asynchronous mode of operation, has demonstrated remarkable reductions in energy costs while avoiding the manufacturing difficulties of analogue equivalents \cite{Merolla:TrueNorth_Science,Akopyan:TrueNorth}.
The spiking neural networks can also be designed using spintronics devices\cite{MTJ_SNN, MTJ_SNN2, MTJ_SNN3}.

A prime example of a versatile neuromorphic architecture leveraging digital spiking neural networks, is the \textit{IBM TrueNorth NS1e} platform \cite{Sawada:NS1e,Cassidy:TrueNorth_Core}. Using a variation of \ac{LIF} neuron models, \textit{TrueNorth NS1e} aims to trade-off between realistically emulating neuronal behaviour and energy efficient hardware, allowing for the execution of large-scale, real-time applications, while limiting power consumption \cite{Merolla:TrueNorth_Science, Izhikevich:SpikingModels}. Recently, the \textit{TrueNorth NS1e} has been applied to solve novel problems, such as implementing a \textit{neuromorphic sieve} algorithm on the \textit{TrueNorth NS1e} architecture to perform efficient integer factorization for cryptographic applications \cite{Monaco:Spiking_Factor, Monaco:NeuroSieve}.

\section{Stochastic Computing}
\label{sec:stochastic}

\begin{figure}[t]
	\centering
	\includegraphics[width=1\linewidth]{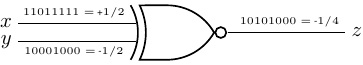}
    \caption{Bipolar stochastic computing multiplier.}
	\label{fig:SC_mult_bipolar}
\end{figure}

Stochastic computation represents information with sequences of random bits \cite{Gaines:Stochastic}. There are two mappings commonly used: unipolar and bipolar  coding. For a sequence of bits $x(t)$, we denote the probability of observing a \enquote{1} to be $P_x = $Pr$(x(t)=1)$. In unipolar coding, the represented value $X$ is $X= P_x, (0 \leq X \leq 1)$. In bipolar coding, the represented value $X$ is $X = (2\cdot P_x -1),  (-1 \leq X \leq 1)$. The stochastic bit streams are generated using a binary to stochastic converter, where the \ac{PRNG} is often realized using a \ac{LFSR}. The input variable, $x$, is then compared with a random number that generates a stochastic bit stream. \cref{fig:SC_mult_bipolar} shows a stochastic two-input multiplier in bipolar coding. The multiplier is in essence a single two-input XNOR gate. The input and output probabilities are represented using $N$-bit length streams, and $N$ clock cycles are required to complete the multiplication.

\begin{figure}[t]
	\centering
	\includegraphics[width=1\linewidth]{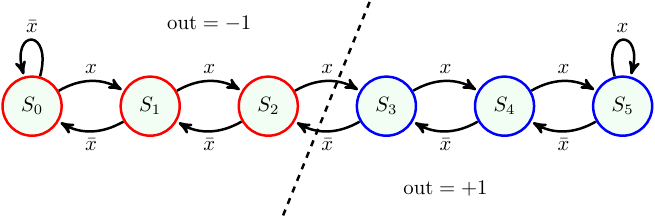}
    \caption{Stochastic \acs{FSM}-based \emph{tanh} function.}
	\label{fig:FSM_tanh}
\end{figure}

\begin{figure}[t]
	\centering
	\includegraphics[width=0.85\linewidth]{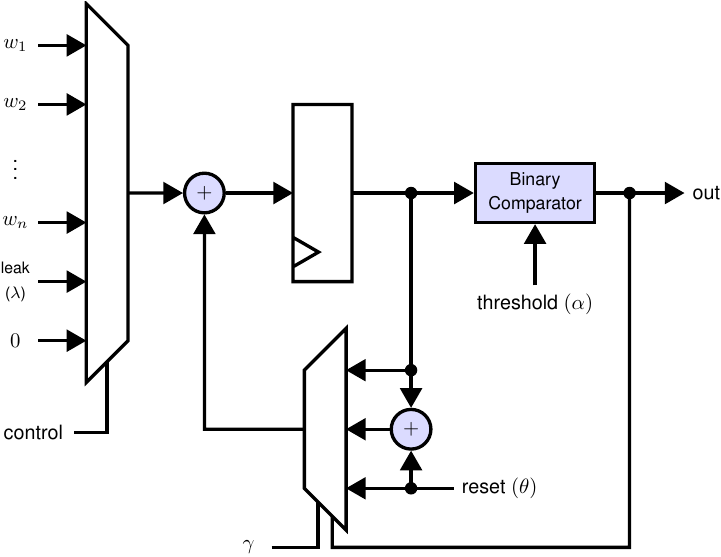}
    \caption{Generalized digital spiking neuron model.}
	\label{fig:spiking_neuron}
\end{figure}

Examining the digital spiking neuron models themselves, it has been demonstrated that their operations are analogous to those of stochastic computing \acp{FSM}. Variations in the programmable parameters of a general spiking neuron can be mapped to various \ac{FSM} structures (such as that shown in \cref{fig:FSM_tanh}), when data is represented as a rate code \cite{Smithson:StochasticSpiking}. For example, when using \acp{FSM} in stochastic computation, hyperbolic tangent functions are simply realized, \cref{fig:FSM_tanh}. In the \ac{FSM}-based functions, the state transitions to the right (next higher state) when the input stochastic bitstream ($x)$ is high, or transitions to the left (next lower state) when low. The output stochastic bitstream ($out$), is determined at each clock cycle by the current state (the output is high for states $S_3$ through $S_5$, and is otherwise low). Individual variations of the programmable parameters, for example those in \cref{fig:spiking_neuron} ($\alpha$, $\gamma$, $\lambda$, etc.) can then be mapped to various stochastic \acp{FSM}, such as that shown in \cref{fig:FSM_tanh} (e.g. number of states, or state transition rules can be adjusted) \cite{Li:FSM,Smithson:StochasticSpiking}. The resulting behaviour of the stochastic \emph{tanh} function, \emph{Stanh}, defined as:
\begin{equation}
    Stanh\big( N_T, x \big) \approx tanh\big(x \cdot N_T/2 \big),
    \label{eqn:tanh}
\end{equation}
where $N_T$ is the total number of states.
Taking advantage of stochastic computing circuits can yield many benefits including simple hardware implementations that are inherently fault tolerant and require little area (low cost) \cite{BrownCard:StochasticNeuralI}. By nature, stochastic computing has a run-time flexibility between execution time and precision. For example, the bipolar stochastic multiplier of \cref{fig:SC_mult_bipolar} can achieve the equivalent of $3$-bits of precision when run for $8$ clock cycles and the hardware remains unchanged even if $4$-bits of precision are required and the system run for $16$ clock cycles.

\section{Proposed Hardware Model}
\label{sec:architecture}

\subsection{Boltzmann Machine Logic Representation}
\label{sec:boltzmann_architecture}

\begin{figure}[t]
	\centering
	\includegraphics[width=0.45\linewidth]{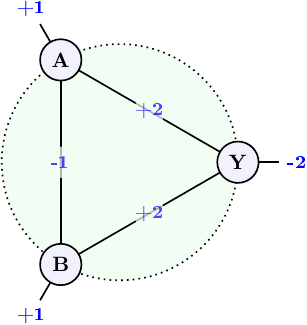}
    \caption{Invertible AND gate Boltzmann machine.}
	\label{fig:inv_and}
\end{figure}

The underlying structure of an instance of the proposed invertible logic circuits can be represented as a network or graph of simple processing elements interconnected to form Boltzmann machine structures \cite{Smithson:StochasticSpiking,Hinton:BoltzmannLearning,Boltzmann1984}. As parallel computational organizations of nodes (simple processing elements) with binary inputs and outputs, Boltzmann machines are well suited for stochastic computing implementations. As illustrated in \cref{fig:inv_and}, every node in the graph is fully-connected to all others through bidirectional links (the weight of the input to node $A$ from node $B$ is equal to the weight of the input to node $B$ from node $A$) and each is assigned an individual bias value.
The outputs of each node can then be computed by taking the weighted sum of all input connections, adding the bias terms to a noise source, and finally applying a non-linear activation function to the sum. The key differences from other common \ac{ANN} topologies are the addition of a noise source and the fact that the outputs are constrained to take on only binary values of $+1$ or $-1$. The ${\rm i^{th}}$ node output behaviour can be written as:
\begin{subequations}
    \begin{equation}
        m_i\left(t+\tau\right) = sgn\Big(rnd\left(-1,+1\right)+tanh\big( I_i\left(t+\tau\right) \big)\Big),
        \label{eqn:boltzmann_output1}
    \end{equation}
    \begin{equation}
        I_i\left(t+\tau\right) = I_0\Big(h_i+\sum_j J_{ij}m_j\left(t\right)\Big),
        \label{eqn:boltzmann_output2}
    \end{equation}
    \label{eqn:boltzmann_output}
\end{subequations}
where $rnd \left(-1, +1 \right)$ is a uniformly distributed random (real) number between $-1$ and $+1$, $sgn$ is the sign function (with binary $+1$ or $-1$ outputs), $I_0$ is a scaling factor (an inverse \textit{pseudo-temperature}), $h$ is the bias vector, and $J$ the weight matrix.

\begin{figure}[t]
    \centering
    \subfloat[Expected states with output clamped to $Y=0$.]{
        \includegraphics[width=0.65\linewidth]{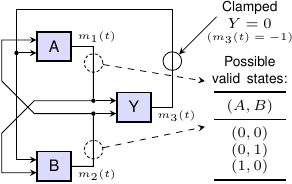}
        \label{fig:inv_and_d}
    }\\
    \subfloat[Expected states with output clamped to $Y=1$.]{
        \includegraphics[width=0.65\linewidth]{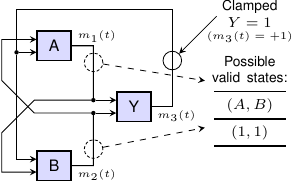}
        \label{fig:inv_and_e}
    }
    \caption{Invertible operations for AND function.}
    \label{fig:inv_AND_ex}
\end{figure}

The Boltzmann machine can be operated by \enquote{clamping} the outputs of any given nodes, and allowing the remaining to be computed. Such a configuration will tend towards the lowest energy states (with connection weights chosen such that the lowest energy states are valid Boolean logic combinations); the random noise in \cref{eqn:boltzmann_output} is included to avoid getting trapped in local minima \cite{Hinton:BoltzmannLearning}. As an illustrative example, consider the invertible AND function of \cref{fig:inv_AND_ex}. There are three nodes, those denoted \emph{A} and \emph{B} refer to binary inputs and \emph{Y}, the output. In invertible logic, we define such nodes as being bidirectionally connected; what is traditionally considered an input may behave as such, and can also be used as output terminals. For a given Hamiltonian defining an invertible gate, possible values for the weights and biases are shown in \cref{eqn:hJ_AND}, with corresponding Boltzmann machine graph in \cref{fig:inv_and}\cite{GSS1,Camsari:pbits_X}.
%
\begin{subequations}
    \begin{equation}
        {\rm h_{AND}} = \left[
        \begin{array}{rrr}
            +1 & +1 & -2
        \end{array}
        \right]
    \end{equation}
    \begin{equation}
        {\rm J_{AND}} = \left[
        \begin{array}{rrr}
             0 & -1 & +2 \\
            -1 &  0 & +2 \\
            +2 & +2 &  0
        \end{array}
        \right]
    \end{equation}
    \label{eqn:hJ_AND}
\end{subequations}
In \cref{eqn:hJ_AND} each $h_i$ is assigned to a node (e.g. $h_1$ for \emph{A}, $h_2$ for \emph{B}, and $h_3$ for \emph{Y}) and each row of $J$ is also assigned to a node. After assigning $h$ and $J$ to the nodes, the states of (\emph{A}, \emph{B}, and \emph{Y}) are categorized to valid and invalid states. Note that a logic value of \enquote{0} is assigned to $m_i(t)=-1$ and \enquote{1} is assigned to $m_i(t)=1$. If the nodes are unconstrained, the system fluctuates among all possible valid states. If one or more of the nodes is clamped to a certain value, the other nodes will fluctuate among all the valid states in which the clamped values are present. For instance, when \emph{Y} is clamped to \enquote{0}, the valid states of (\emph{A}, \emph{B}) are ($0$, $0$), ($0$, $1$), and ($1$, $0$), since those are the input combinations which cause the output of an AND-gate to be equal to \enquote{0}, as shown in \cref{fig:inv_and_d}. Likewise, when \emph{Y} is clamped to \enquote{1}, the only valid state of (\emph{A}, \emph{B}) is ($1$, $1$) as shown in \cref{fig:inv_and_e}. The tendency of the system to fluctuate between valid states given a clamped output makes it possible to operate the gate in reverse to compute the original inputs.

\subsection{Base Processing Element Architecture}
\label{sec:neuron_architecture}

\begin{figure}[t]
	\centering
	\includegraphics[width=1\linewidth]{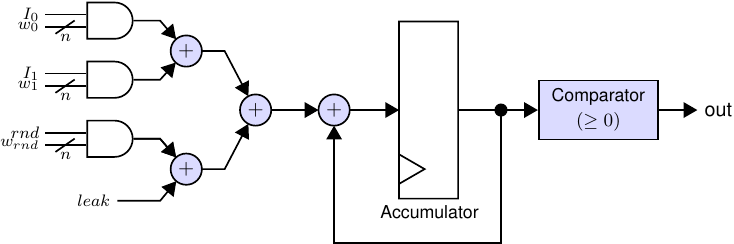}
    \caption{Base processing element using stochastic computing (simplified spiking neuron in rate coding) model.}
	\label{fig:spiking_pbit}
\end{figure}

Node behaviour, as described by \cref{eqn:boltzmann_output}, is nearly in the ideal form for the use of a stochastic \ac{FSM} \cite{Li:FSM}; the outputs of each node are stochastic bitstreams varying between $-1$ and $+1$, well suited for bipolar coding \cite{Gaines:Stochastic,Poppelbaum:Stochastic}. The hyperbolic-tangent function is easily implemented with such an \ac{FSM}. Implementing the transfer function as shown in \cref{eqn:boltzmann_output} without modification, would require the hyperbolic-tangent function to be implemented with a stochastic \ac{FSM} as shown in \cref{fig:FSM_tanh}. Furthermore, the output bitstream would have to be summed with a randomly generated signal; a task which is non-trivial using stochastic addition circuitry \cite{Gaines:Stochastic}. An alternative is to perform the summation of the weighted noise source (with corresponding weight denoted as $w_{rnd}$) alongside the computation of the weighted sum of input signals, transforming \cref{eqn:boltzmann_output} into:
\begin{subequations}
    \begin{equation}
        m_i\left(t+\tau\right) = sgn\Big(tanh\big( I_i\left(t+\tau\right) \big)\Big),
        \label{eqn:boltzmann_modified1}
    \end{equation}
    \begin{equation}
        I_i\left(t+\tau\right) = h_i+\sum_j J_{ij}m_j\left(t\right) + w_{rnd} \cdot sgn\Big(rnd\big(-1,+1\big)\Big),
        \label{eqn:boltzmann_modified2}
    \end{equation}
    \label{eqn:boltzmann_modified}
\end{subequations}
where the $I_0$ scaling value in \cref{eqn:boltzmann_output} can be included in the $h$ and $J$ terms, and is omitted in \cref{eqn:boltzmann_modified}. 
The characterized behaviour is implemented using the base processing element model using stochastic computing (spiking neuron model in rate coding) shown in \cref{fig:spiking_pbit}, which is a simplified implementation of \cref{fig:spiking_neuron}\cite{Merolla:TrueNorth_Science, Smithson:StochasticSpiking}; a single stochastic input signal is added with a fixed weight as the noise source. 
The model can be reduced through the removal of the programmable threshold ($\alpha$) and reset behaviours ($\gamma$ and $\theta$) from \cref{fig:spiking_neuron}. Furthermore, due to the few inputs to each node, computing the input summation can be simplified and the final configuration of \cref{fig:spiking_pbit} obtained (where the $rnd$ node in \cref{fig:spiking_pbit} represents a single-bit random signal). For all the Boltzmann machine implementations in this work (e.g. as shown in \cref{fig:inv_and}), each node was implemented using the model in \cref{fig:spiking_pbit}.
During operation, a Boltzmann machine will settle into a local energy minimum if left running freely with no input. However, if the amplitude of the random noise source (referred to as $w_{rnd}$) is incrementally reduced, then the resulting behaviour is analogous to that of \emph{simulated annealing} where the node outputs will initially highly stochastic components, and eventually settle at a low energy state corresponding to a desired circuit output value \cite{Hinton:BoltzmannLearning,Kirkpatrick:SimulatedAnnealing}.

\subsection{Pseudo-Random Number Generation}
\label{sec:architecture_prng}

At the core of the neuron model we used is a saturating accumulator with behaviour that can be observed in the graphical representation of the equivalent stochastic \ac{FSM} shown in \cref{fig:FSM_tanh}. The \ac{FSM} state is stored as the accumulator, and once at the highest state ($S_5$), a positive input ($x$) will not cause any increase of the value stored in the accumulator. Likewise, the inverse holds true for negative input signals ($\bar{x}$) when at the lowest state ($S_0$). Ideally, the accumulator (and subsequent adders in \cref{fig:spiking_pbit}) should be of the lowest precision required for a given $J$ and $h$ pair. However, if the input and output values are averaged over many clock cycles, then the presence of DC bias in the generated pseudo-random bitstream $\left(\text{i.e. } \lim_{t \to \infty} \left( \sum_{\tau=0}^{t} {rnd \left(\tau \right)}/{t} \right) \neq 0.5 \right)$ can lead to the saturation of system states independent of neuron input signals, leading to increases in the number of cycles the system spends in invalid states. In our experiments, simple \ac{LFSR} circuits resulted in much higher output errors (time spent at invalid states) compared to larger (more complicated and costly) pseudo-random number generation schemes. For the simulation results presented in this paper, all random bitstreams were generated by $64$-bit \textit{xorshift+} circuits, with each bit of the register feeding a single neuron element \cite{Vigna:xorshift}.

\section{Invertible Logic Circuits}
\label{sec:inv_logic_circuits}

Boltzmann machine configurations for all basic gates (three-terminal AND, NAND, OR, NOR, etc.) can be derived following the same steps as those to design the AND gate structure of \cref{fig:inv_and}. However, for more complex structures, auxiliary bits may have to be added.

\subsection{Invertible Binary Adder Circuits}
\label{sec:inv_adders}

\begin{figure}[t]
    \centering
    \subfloat[Invertible XOR (\& OR).]{
        \includegraphics[width=0.375\linewidth]{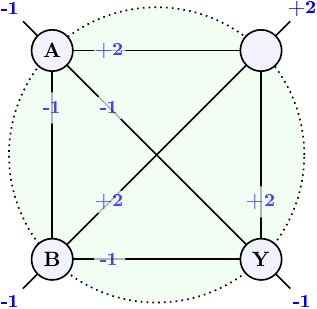}
        \label{fig:inv_xor_or}
    }
    \hspace{0.005\columnwidth}
    \subfloat[Invertible XOR (\& NOR).]{
        \includegraphics[width=0.375\linewidth]{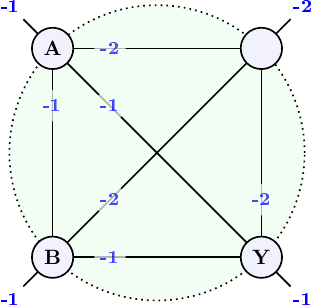}
        \label{fig:inv_xor_nor}
    }\\
    \subfloat[Invertible half-adder.]{
        \includegraphics[width=0.375\linewidth]{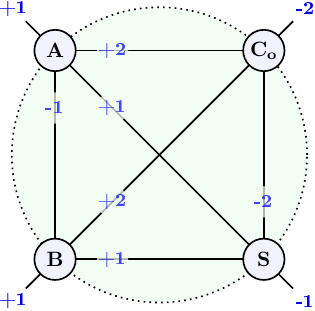}
        \label{fig:inv_ha}
    }
    \caption{Boltzmann machine configurations of XOR gates.}
    \label{fig:inv_xors}
\end{figure}

\begin{figure}[t]
	\centering
    \subfloat[Alternate invertible half-adder.]{
        \includegraphics[width=0.45\linewidth]{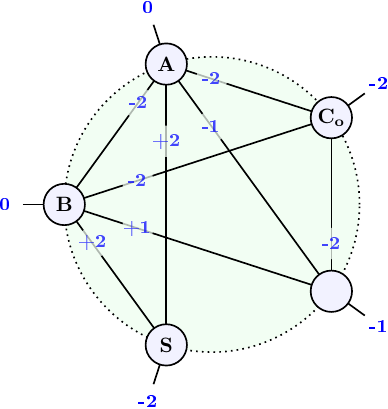}
		\label{fig:inv_ha_alt}
	}
    \subfloat[Invertible full-adder.]{
        \includegraphics[width=0.45\linewidth]{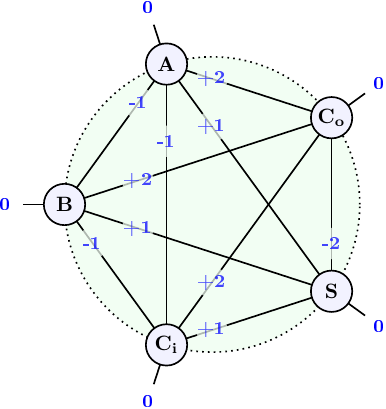}
	    \label{fig:inv_fa}
    }\\
    \subfloat[Combined invertible half-adder with AND gate.]{
        \includegraphics[width=0.45\linewidth]{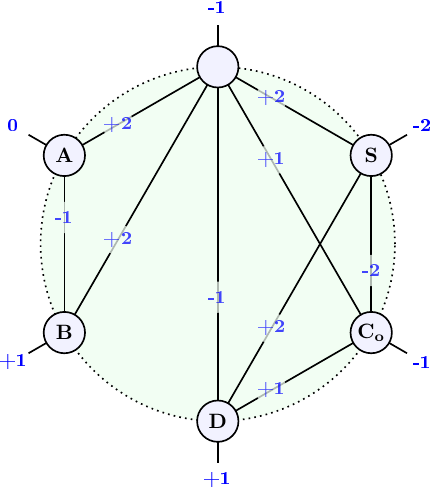}
        \label{fig:ha_and}
    }
    \caption{Invertible adder Boltzmann machines.}
	\label{fig:inv_half-adders}
\end{figure}

XOR gate implementations require the addition of only a single auxiliary bit, as shown in \cref{fig:inv_xors}. However, these auxiliary nodes can instead be used as additional outputs through careful choice of network weights. For example, \cref{fig:inv_xor_or} is designed such that the spare node outputting the Boolean function $A\lor B$, while \cref{fig:inv_xor_nor} results in $\overline{A\lor B}$ instead; in both circuits, the output remains the desired $Y = A \oplus B$.
%
\begin{subequations}
    \begin{equation} 
        {\rm h_{XOR_{OR}}} = \left[
        \begin{array}{rrrr}
            -1 & -1 & -1 & +2
        \end{array}
        \right]
    \end{equation}
    \begin{equation}
        {\rm J_{XOR_{OR}}} = \left[
        \begin{array}{rrrr}
             0 & -1 & -1 & +2 \\
            -1 &  0 & -1 & +2 \\
            -1 & -1 &  0 & +2 \\
            +2 & +2 & +2 &  0
        \end{array}
        \right]
    \end{equation}
    \label{eqn:hJ_XOR_OR}
\end{subequations}
%
\begin{subequations}
    \begin{equation} 
        {\rm h_{XOR_{NOR}}} = \left[
        \begin{array}{rrrr}
            -1 & -1 & -1 & -2
        \end{array}
        \right]
    \end{equation}
    \begin{equation}
        {\rm J_{XOR_{NOR}}} = \left[
        \begin{array}{rrrr}
             0 & -1 & -1 & -2 \\
            -1 &  0 & -1 & -2 \\
            -1 & -1 &  0 & -2 \\
            -2 & -2 & -2 &  0
        \end{array}
        \right]
    \end{equation}
    \label{eqn:hJ_XOR_NOR}
\end{subequations}
This characteristic of XOR (which also applies to XNOR and other similar gates) allows for a very compact binary half-adder circuit to be designed. The structure characterized by \cref{fig:inv_ha} and \cref{eqn:hJ_HA} is composed of an invertible XOR gate designed with an auxiliary AND output. This results in a much more compact structure than that of \cref{fig:inv_ha_alt} which can be derived through the combination of the Hamiltonian matrices of separate invertible AND and XOR gates \cite{GSS2}. Subsequently, we also designed a Boltzmann machine implementation of a full-adder of the most compact form (a network of $5$ nodes), as shown in \cref{fig:inv_fa} and \cref{eqn:hJ_FA}. Once again, the network designed functionally implements an invertible full-adder with much fewer nodes than required by existing methods \cite{GSS2}.
%
\begin{subequations}
    \begin{equation} 
        {\rm h_{HA}} = \left[
        \begin{array}{rrrr}
            +1 & +1 & -1 & -2
        \end{array}
        \right]
    \end{equation}
    \begin{equation}
        {\rm J_{HA}} = \left[
        \begin{array}{rrrr}
             0 & -1 & +1 & +2 \\
            -1 &  0 & +1 & +2 \\
            +1 & +1 &  0 & -2 \\
            +2 & +2 & -2 &  0
        \end{array}
        \right]
    \end{equation}
    \label{eqn:hJ_HA}
\end{subequations}
%
\begin{subequations}
    \begin{equation} 
        {\rm h_{HA_2}} = \left[
        \begin{array}{rrrrr}
             0 &  0 & -2 & -1 & -2
        \end{array}
        \right]
    \end{equation}
    \begin{equation}
        {\rm J_{HA_2}} = \left[
        \begin{array}{rrrrr}
             0 & -2 & +2 & -1 & -2 \\
            -2 &  0 & +2 & +1 & -2 \\
            +2 & +2 &  0 &  0 &  0 \\
            -1 & -1 &  0 &  0 & -2 \\
            -2 & -2 &  0 & -2 &  0
        \end{array}
        \right]
    \end{equation}
    \label{eqn:hJ_HA_alt}
\end{subequations}
%
\begin{subequations}
    \begin{equation} 
        {\rm h_{FA}} = \left[
        \begin{array}{rrrrr}
            0 & 0 & 0 & 0 & 0
        \end{array}
        \right]
    \end{equation}
    \begin{equation}
        {\rm J_{FA}} = \left[
        \begin{array}{rrrrr}
             0 & -1 & -1 & +1 & +2 \\
            -1 &  0 & -1 & +1 & +2 \\
            -1 & -1 &  0 & +1 & +2 \\
            +1 & +1 & +1 &  0 & -2 \\
            +2 & +2 & +2 & -2 &  0
        \end{array}
        \right]
    \end{equation}
    \label{eqn:hJ_FA}
\end{subequations}

\subsection{Invertible Binary Multiplier Circuits}
\label{sec:inv_multipliers}

\begin{figure}[t]
	\centering
    \subfloat[$4$-bit invertible multiplier configuration.]{
        \includegraphics[width=0.8\columnwidth]{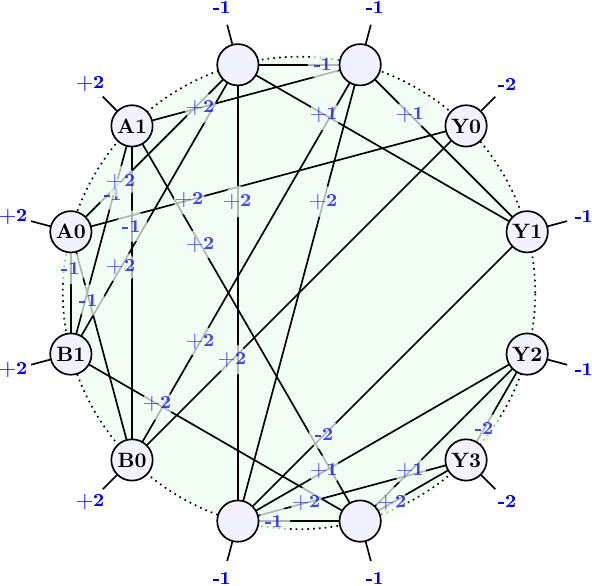}
        \label{fig:2bit_mult}
    }\\
    \subfloat[$6$-bit invertible multiplier configuration.]{
        \includegraphics[width=1\columnwidth]{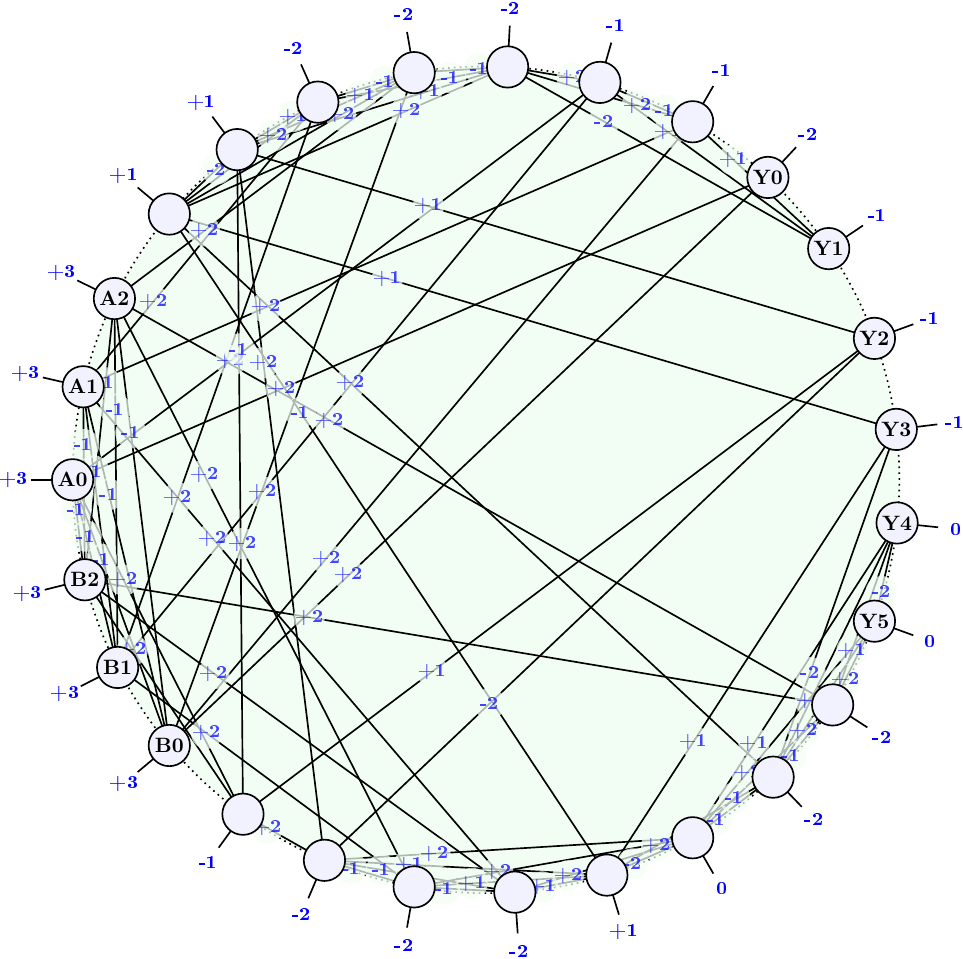}
        \label{fig:3bit_mult}
    }
    \label{fig:boltzmann_machines}
    \caption{Invertible multiplier Boltzmann machine configurations.}
\end{figure}

Ultimately, the goal of designing compact invertible adder circuits (as well as those of basic gates) is for use as building blocks for designing more complex, yet still fully invertible, logic circuits. In order to demonstrate this, we designed invertible (unsigned) multiplier circuits by combining the structure of invertible full-adders, half-adders, and AND gates. When combining multiple invertible logic circuits, such as when an AND gate is combined with a half-adder in \cref{fig:ha_and} in order to obtain $\left(A \land B \right) + D$, the output node of the AND gate can be fused with one of the input nodes of the half-adder. The results of such operations may yield sparsely connected graphs with irregularity in the number of inputs to each node, however there is no guarantee that denser, more compact structures do not exist. The resulting graphical representation of a $2$-bit by $2$-bit multiplier (with $4$-bit output) is shown in \cref{fig:2bit_mult}, and the similar structure for a multiplier circuit with a $6$-bit output is shown in \cref{fig:3bit_mult}. Larger multiplier circuits become difficult to visualize in two-dimensional space, but are designed in much the same manner.

\section{Results}
\label{sec:experimental_results}

In order to evaluate the proposed invertible logic circuits, fully-synthesizable SystemVerilog designs were first simulated using a SystemC environment. In all cases, the pseudo-random noise sources were generated in hardware (not software). Furthermore, all circuit inputs were fixed logic values (no stochastic signals are required to be used as inputs, simplifying application circuit designs) and the outputs reported are the statistics of the output states measured over the specified number of clock cycles for each experiment. The metric used to evaluate circuit accuracy is defined as the percentage of clock cycles spent in valid states, and where a valid state is any which satisfies the desired logical expression being evaluated. In addition to simulation results, fabricated \ac{ASIC} test results and measurements are also provided, alongside \ac{FPGA} synthesis results for a variety of invertible multiplier configurations. For all hardware implementations, the hardware costs (in terms or circuit area for the fabricated \ac{ASIC} and required \acp{LUT} and registers for \ac{FPGA} results) are presented and compared to existing works when possible.

\subsection{Multiplication Results}
\label{sec:multiplication_results}

So as to demonstrate the correct operation of the proposed invertible multiplier, the behaviour of an $8$-bit multiplier (calculating the $8$-bit product of two $4$-bit inputs) was first analyzed. The circuit was realized as a scaled equivalent to the $4$-bit multiplier illustrated in \cref{fig:2bit_mult}; in total, $48$ spiking neurons were required, and a $64$-bit \textit{xorshift+} \ac{PRNG} circuit was used. Simulations were performed with all neurons using $4$-bit weights. 

\begin{figure}[t]
	\centering
    \subfloat[Histogram plot of invertible multiplier output.]{
        \includegraphics[width=0.9\linewidth]{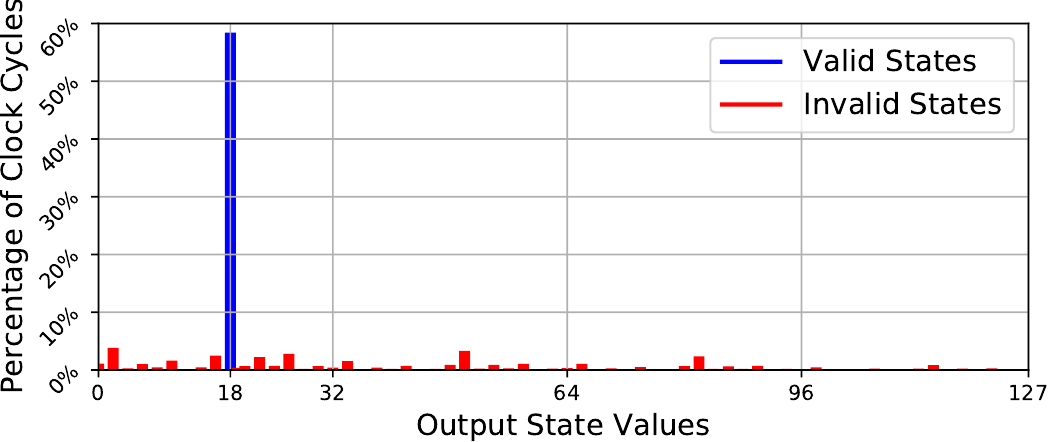}
		\label{fig:mult_bar}
	}\\
	\subfloat[Invertible multiplier output state convergence.]{
        \includegraphics[width=0.9\linewidth]{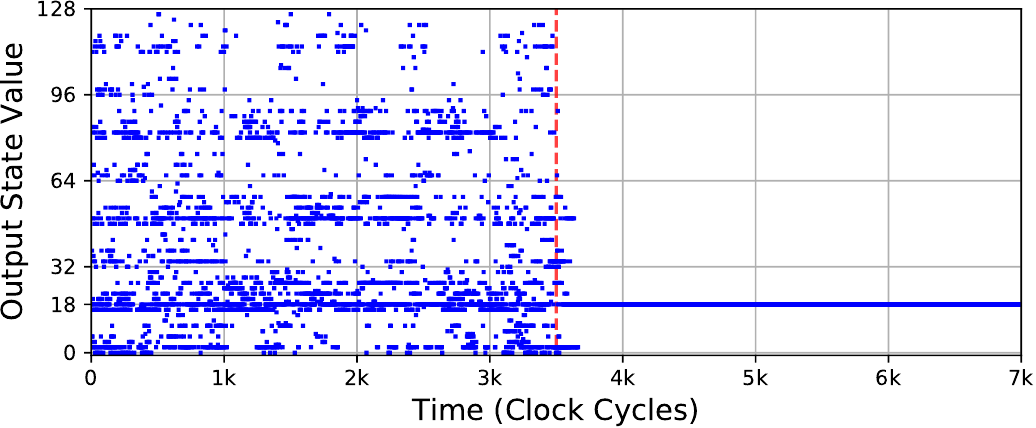}
		\label{fig:mult_time}
    }\\
    \subfloat[Invertible multiplier output state convergence (enlarged).]{
        \includegraphics[width=0.9\linewidth]{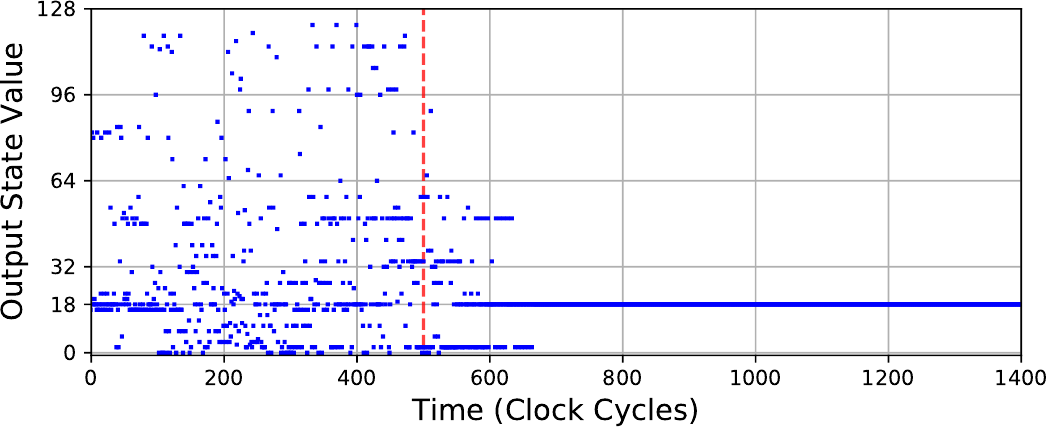}
		\label{fig:mult_time_zoom}
	}
    \caption{Invertible multiplier simulation results (for input pair $3$x$6$).}
	\label{fig:mult_results}
\end{figure}

\begin{figure}[t]
    \centering
    \includegraphics[width=0.9\linewidth]{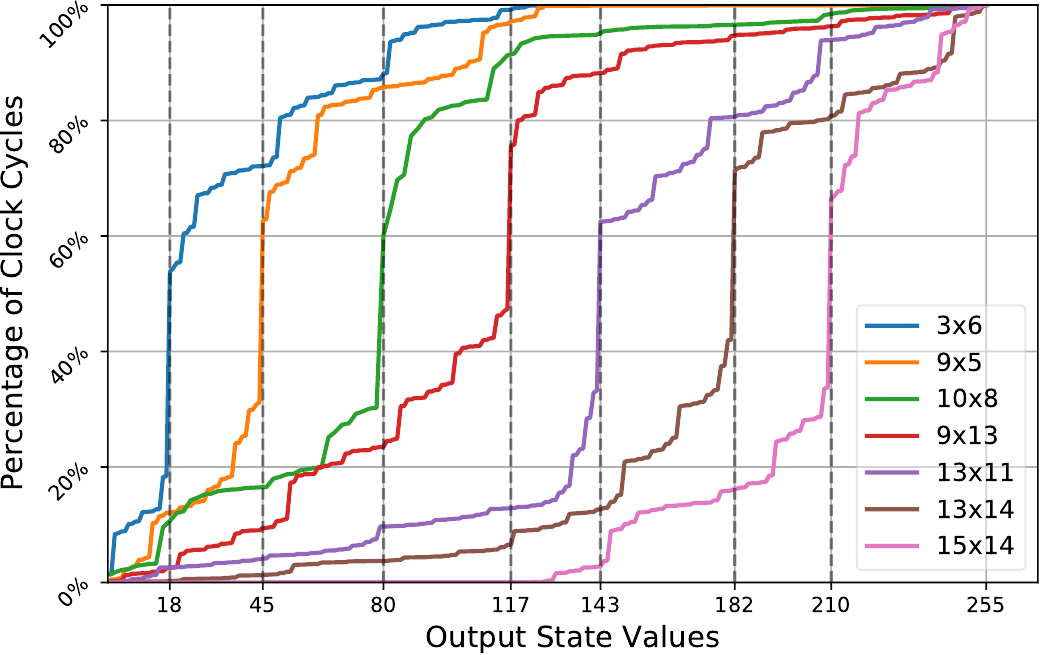}
    \caption{\acs{CDF} plots of invertible multiplier outputs.}
    \label{fig:mult_cdf}
\end{figure}

\cref{fig:mult_bar} plots the histogram of the output state distribution of the simulated circuit operating as a multiplier. In order to plot stable mean state values, the simulations were run for an extended period of $N=2^{20}$ clock cycles, and the weight of all random signals ($w_{rnd}$ in \cref{eqn:boltzmann_modified2}) was decreased from an initial value of $w_{rnd}=5$ to $w_{rnd}=3$ at time $t=N/2$; the time domain behaviour of the output state values centred around $t=N/2$ is plotted in \cref{fig:mult_time}. The aforementioned values of $w_{rnd}$ chosen were experimentally obtained by first running the invertible logic circuit with fixed noise amplitudes and measuring the output switching activity; the final values selected were those yielding a balance between initially noisy outputs and a stable output value. Future work will explore measuring, or estimating, the Boltzmann machine energy state during circuit operation in order to automatically adjust the noise input patterns accordingly. It should be noted that the time domain plot in \cref{fig:mult_time} are run for much longer time periods than required (as evident by the stable output values after $w_{rnd}$ is reduced) and the lengthy durations are included for readability; an enlarged plot is also included in \cref{fig:mult_time_zoom}. Evident in \cref{fig:mult_time} is that before the decrease in the random bitstream weights, the output state fluctuated among many possible values, and after $t=N/2$ the output converged to the single valid state (output value of $18$ for the tested input combination of $3\rm{x}6$). As such, the results demonstrate that the invertible logic circuit can be used to obtain a valid input combination from only specifying the output value.
The invertible multiplier circuit can also be run with a fixed weighting of the random signals; differing from the results presented in \cref{fig:mult_time}, recovering the computed output values then requires calculating the mode of the output states (a non-trivial task). However, this does demonstrate the flexibility of the method; for designs with multiple valid output values, the overhead of calculating the most common states may be acceptable for some applications. When operating with a fixed random weight of $w_{rnd}=5$, the same multiplier circuit was simulated for various input value combinations, and the \acp{CDF} of the output state values plotted in \cref{fig:mult_cdf}. As was performed for the previous experiments, the value of $w_{rnd}$ was chosen to obtain an output waveform with enough randomness so as to avoid converging to incorrect output values, while remaining low enough to prevent the output signals from fluctuating in a purely stochastic manner. The final value of $w_{rnd}$ was obtained through experimentally sweeping the random noise source weight and selecting the desired parameter value. From the graph shown in \cref{fig:mult_cdf} it can be seen that in all cases, the mode output states (where the largest increases occur) were for valid output values. In such cases where the outputs were left free-running with constant random noise introduced, the correct output value can be determined simply by taking the mode of the output distribution. The forwards multiplication results presented have mode frequencies in the range of $30\%$ to $40\%$, while the next most frequent states are only measured to be present for $\approx 5\%$ of the clock cycles; a significant discrepancy, which allows for the desired output values to be easily distinguished from invalid states.

\subsection{Factorization Results}
\label{sec:factorization_results}

\begin{figure}[ht]
	\centering
    \subfloat[Histogram plot of invertible factorizer inputs.]{
        \includegraphics[width=0.9\linewidth]{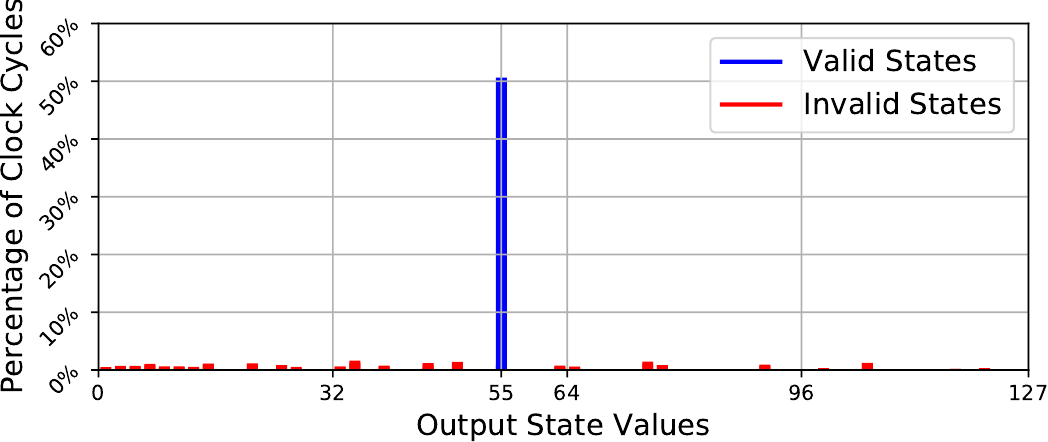}
		\label{fig:fact_bar}
	}\\
	\subfloat[Invertible factorizer input states convergence.]{
        \includegraphics[width=0.9\linewidth]{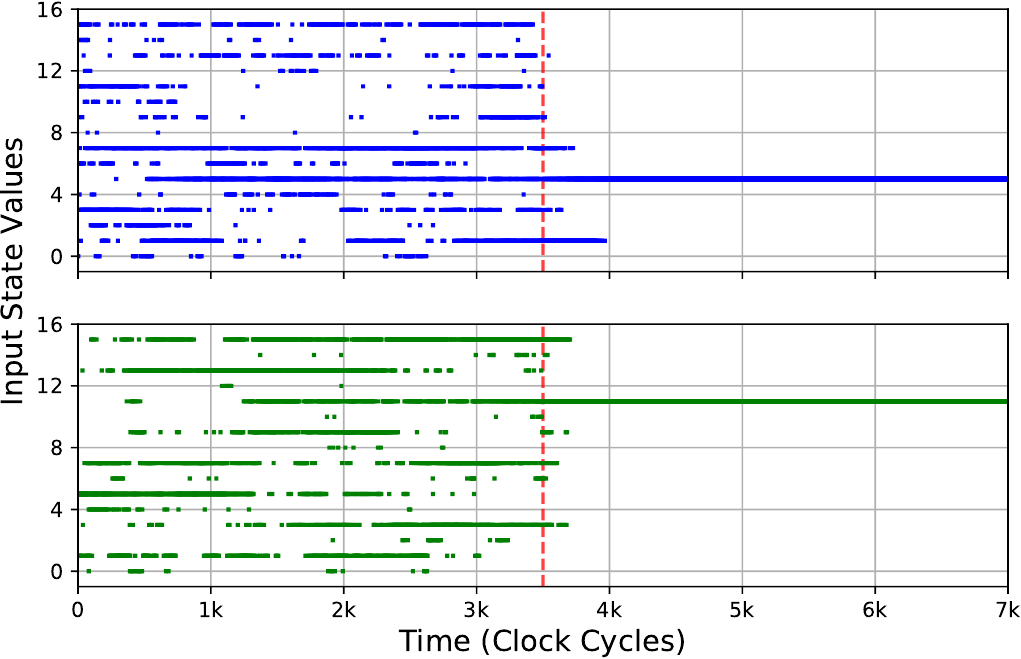}
		\label{fig:fact_time}
    }\\
    \subfloat[Invertible factorizer input states convergence (enlarged).]{
        \includegraphics[width=0.9\linewidth]{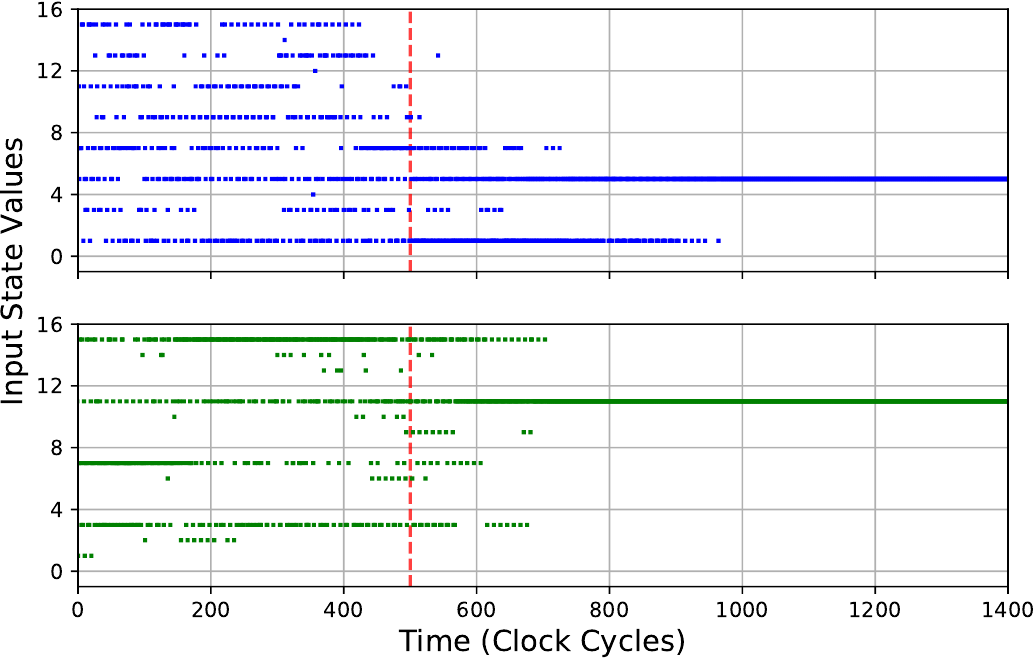}
		\label{fig:fact_time_zoom}
	}
    \caption{Invertible factorizer simulation results (for output value of $55$).}
	\label{fig:fact_results}
\end{figure}

Mirroring the multiplier results of \cref{sec:multiplication_results}, the same circuit was evaluated when operating as a factorizer. All experiments were performed with neurons using $5$-bit weights, and the random weights were reduced from initial values of $w_{rnd}=11$ at $t=N/2$ to $w_{rnd}=5$. As with the multiplication results presented above, the resulting values of $w_{rnd}$ chosen were manually selected after having first experimentally swept the parameter over the possible input range. With the output clamped to fixed value of $55$, the histogram in \cref{fig:fact_bar} plots the product of the input states; evident is that the mode states are clearly two with the desired product as output.

Likewise, the time domain results in \cref{fig:fact_time} reiterate the correct operation of the factorizer. 
The figure is the time-domain plots of the two input values of the invertible multiplier (operating as a factorizer).
The state initially fluctuates with time between many possible input values. 
After the noise source amplitude is reduced, the inputs then converge to a single value each. 
The plots are of the numerical values that the inputs take and not their binary values for space considerations.
Evident from the plots is that states of both inputs quickly converge to valid input values of $5$ and $11$ after a short period (input states remain constant after $\approx 500$ cycles). Once again, it should be noted that the time domain plots in \cref{fig:fact_time} are run for much longer time periods than required, for better readability; enlarged plots are also included in \cref{fig:fact_time_zoom}. The exact number of operating cycles required for convergence for the \ac{ASIC} measurement results are tabulated in \cref{table:test_results} in the form of both the mean and worst-case cycles (total times from $t=0$ to output convergence).

\subsection{Synthesis Results}
\label{sec:synthesis_results}

\cref{table:synthesis_results} lists the synthesis results for various sized invertible multipliers using the Synopsis Design Compiler targeting the \ac{TSMC} 65nm \ac{GP} process. Furthermore, the results shown in \cref{table:synthesis_results} illustrate the scalability of the method, with the total area of the invertible multipliers growing at the square of the input bit-width. As a binary array multiplier will grow in area at rate equal to the square of the multiplier bit-width, the results of \cref{table:synthesis_results} demonstrate that the presented method for invertible logic circuits scales at an equal rate.
%
\begin{table}[t]
    \caption{Invertible Multiplier Synthesis Results}
    \label{table:synthesis_results}
    \centering
    \begin{tabular}{C{1.2cm}C{1.2cm}C{1.2cm}C{1.1cm}C{1.1cm}}
        \toprule
        Input bit width & Total area $\left( {\mu \text{m}}^2 \right)$ & Nodes needed & Node area & \acs{PRNG} area \\
        \midrule
        $2$-bit & $19290$ & $12$ & $81.4\%$ & $18.6\%$ \\
        \midrule
        $3$-bit & $48839$ & $27$ & $91.2\%$ & $8.8\%$ \\
        \midrule
        $4$-bit & $92353$ & $48$ & $94.3\%$ & $5.7\%$ \\
        \midrule
        $5$-bit & $153181$ & $75$ & $93.5\%$ & $6.5\%$ \\
        \bottomrule
    \end{tabular}
\end{table}

When targeting programmable \ac{FPGA} fabric, we can directly compare the resources required for our method to that of previous works \cite{Pervaiz:pbits_FPGA}. The number of required \acp{LUT} and registers (when targeting the same Xilinx Kintex Ultrascale XCKU040-1FBVA676) \ac{FPGA} are summarized in \cref{table:FPGA_comparison}. Our work requires significantly less resources for anything more complex than a single invertible AND gate. The sole reason why the presented work requires more \acp{LUT} and registers for a single AND gate is due to a $64$-bit \textit{xorshift+} \ac{PRNG} circuit being instantiated when only three output bits are used. As the structures grow and require more nodes in the Boltzmann machine graphs, this overhead is amortized; the most extreme example is that of a $32$-bit \ac{RCA}, where our work requires less than $27\%$ and $11\%$ of the \acp{LUT} and registers, respectively, than the related work. Furthermore, the more compact Boltzmann machine representation of the \acl{FA} presented in \cref{fig:inv_fa} and \cref{eqn:hJ_FA} results in fewer than $36\%$ of the nodes being required for a single full-adder circuit when compared to previous methods \cite{Pervaiz:pbits_FPGA}. The methodology and circuit constructs presented in this work result in invertible logic structures which are much more compact and cost efficient than any existing work in the field of invertible logic circuits.

\begin{table*}[t]
    \caption{\acs{FPGA} Synthesis Results (Xilinx Kintex Ultrascale XCKU040-1FBVA676)}
    \label{table:FPGA_comparison}
    \centering
    \begin{tabular}{cccccccccc}
        \toprule
        & \multicolumn{3}{c}{Proposed (this work)} & \multicolumn{3}{c}{Conventional \cite{Pervaiz:pbits_FPGA}} & \multicolumn{3}{c}{Proposed/Conventional} \\
        & Nodes & \acsp{LUT} & Registers & Nodes & \acsp{LUT} & Registers & Nodes & \acsp{LUT} & Registers \\
        \midrule
        AND Gate & $3$ & $257$ & $307$ & $3$ & $156$ & $123$ & $100\%$ & $165\%$ & $250\%$ \\
        \midrule
        Full-Adder & $5$ & $400$ & $329$ & $14$ & $1345$ & $586$ & $36\%$ & $30\%$ & $56\%$ \\
        \midrule
        $32$-bit \acs{RCA} & $128$ & $10455$ & $1910$ & $434$ & $38814$ & $18071$ & $29\%$ & $27\%$ & $11\%$ \\
        \bottomrule
    \end{tabular}
\end{table*}

In order to evaluate the overhead of the proposed invertible multiplier/divider/factorizer circuit, the synthesized results are compared to the combined area required for conventional binary multiplier, divider, and trial division factorizing circuits (targeting the same TSMC 65nm GP process). For the largest considered case of a $5$-bit by $5$-bit multiplier, the initial architecture design resulted in the invertible logic circuit requiring $37x$ the area of the conventional binary circuit ($4143\mu\text{m}^2$). However, after manually optimizing the circuit architecture, it was possible to reduce the initial area of $153181\mu\text{m}^2$ by $65\%$, down to $53818\mu\text{m}^2$; reducing the circuit area of the optimized invertible multiplier/divider/factorizer to only $13x$ that of the conventional binary logic counterpart.

In terms of scalability of number of cycles for convergences, the number of cycles to converge to the correct answer is not equal to the number of potential states.
For example, the number of possible states is $2^{27}$ in the 6-bit (3x3) invertible multiplier, while the number of cycles is several hundreds.
In addition, the number of possible states will be less in operation due to the input or output nodes (when being operated in either the forwards of reverse ways) being fixed.  

\subsection{Fabricated Hardware}
\label{sec:hardware}

\begin{figure}[t]
    \centering
    \includegraphics[width=0.85\linewidth]{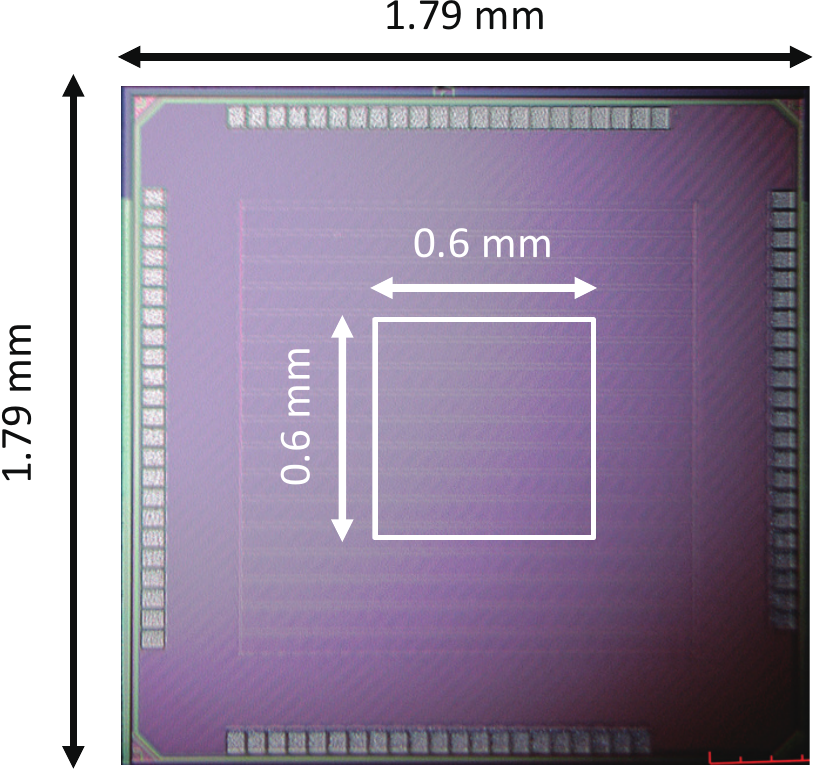}
    \caption{Photomicrograph of fabricated $5$-bit by $5$-bit factorizer circuit.}
    \label{fig:photomicrograph}
\end{figure}

\begin{figure}[ht]
	\centering
	\subfloat[Histogram of convergence cycles in general factorization.]{
		\includegraphics[width=1.0\linewidth]{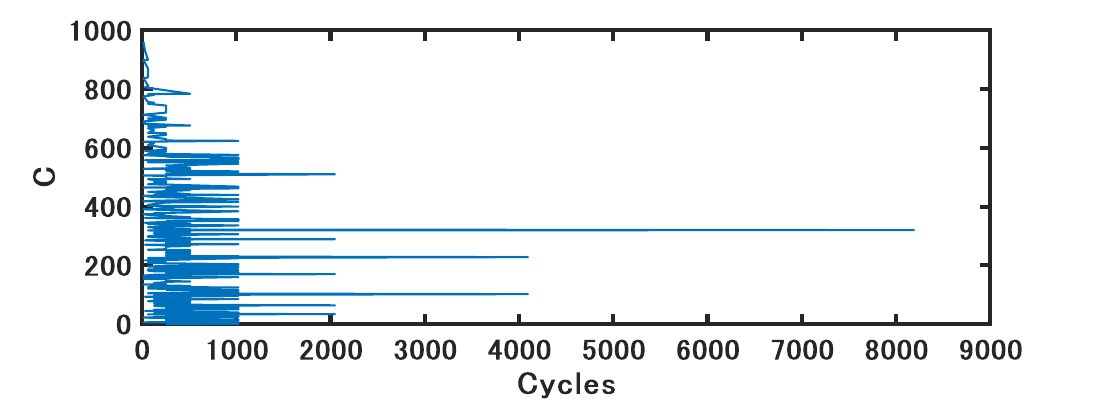}
		\label{fig:hist_all}
	}\\
	\subfloat[Histogram of convergence cycles in prime factorization.]{
		\includegraphics[width=1.0\linewidth]{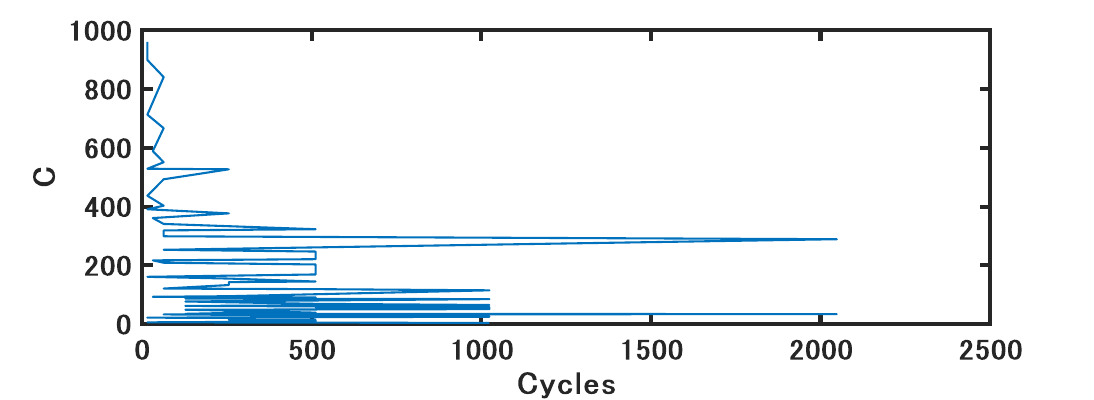}
		\label{fig:hist_prime}
	}
	\caption{Histograms of convergence cycles in the 5-bit by 5-bit factorizer ($A \times B = C$).}
	\label{fig:hardware_hist}
\end{figure}

\begin{figure}[ht]
	\centering
    \subfloat[Invertible factorizer input states convergence ($A \times B = 49$).]{
        \includegraphics[width=0.445\linewidth]{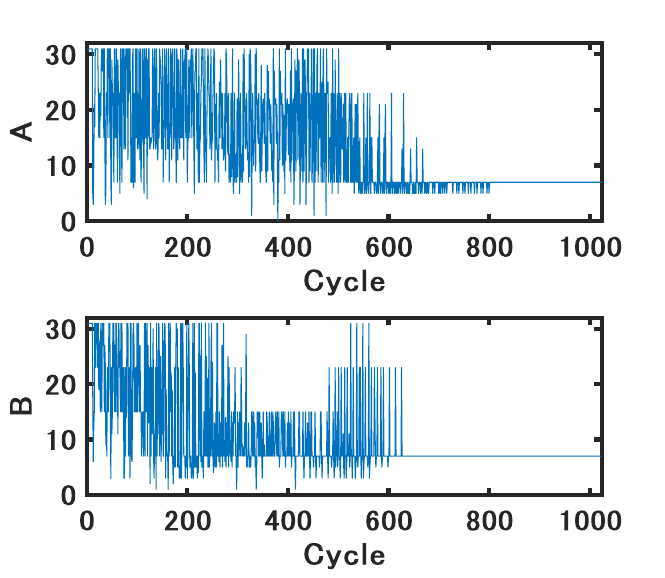}
        \label{fig:hardware_49}
    }\hspace{1mm}
	\subfloat[Invertible factorizer input states convergence ($A \times B = 182$).]{
        \includegraphics[width=0.445\linewidth]{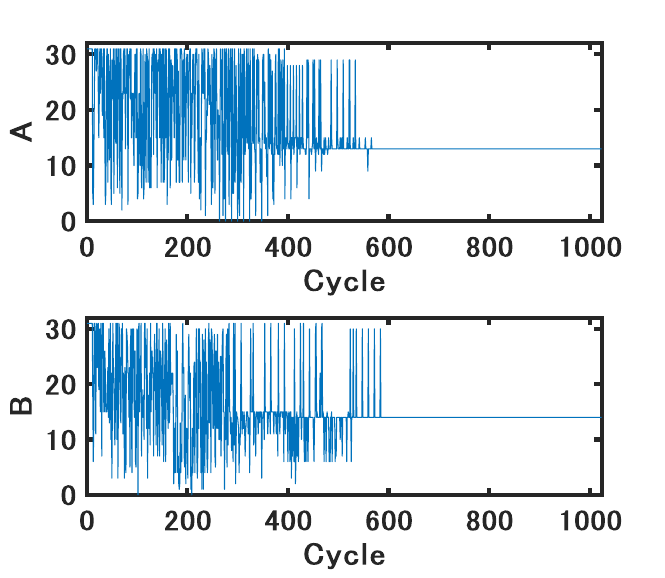}
        \label{fig:hardware_182}
	}\\
	\subfloat[Invertible factorizer input states convergence ($A \times B = 310$).]{
        \includegraphics[width=0.445\linewidth]{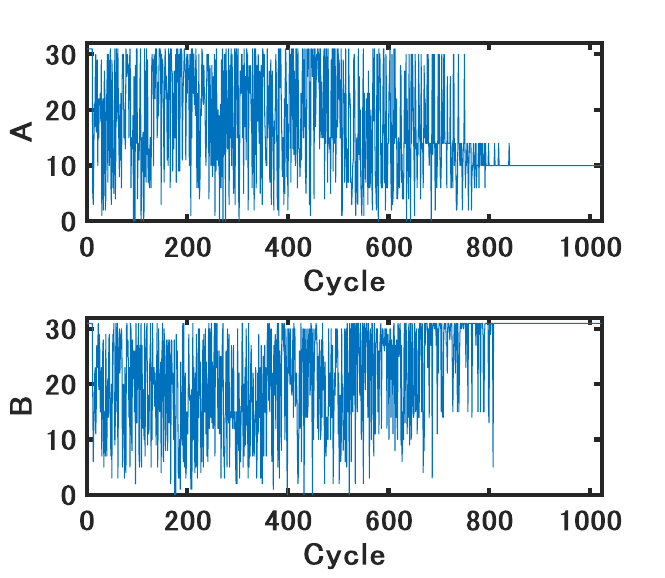}
        \label{fig:hardware_310}
	}\hspace{1mm}
    \subfloat[Invertible factorizer input states convergence ($A \times B = 598$).]{
        \includegraphics[width=0.445\linewidth]{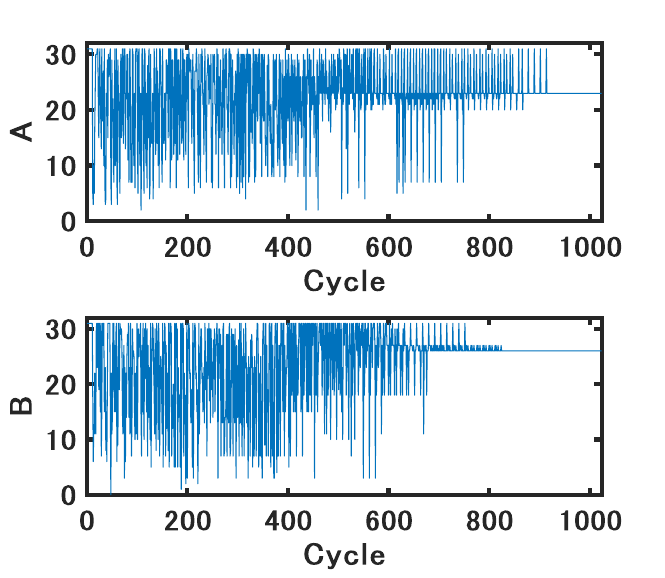}
        \label{fig:hardware_598}
	}
    \caption{Invertible factorizer measured \acs{ASIC} results selected arbitrarily.}
	\label{fig:hardware_results}
\end{figure}

The photomicrograph of a $5$-bit by $5$-bit invertible multiplier is shown in \cref{fig:photomicrograph} with measurement results detailed in \cref{table:test_results}. 
The test chip was designed using SystemVerilog and synthesized using Synopsys Design Compiler in TSMC 65 nm CMOS. 
The layout was obtained using Cadence Innovus for the chip fabrication.
The tabulated test results present the mean, and worst-case values of operating cycles, latency, and resulting energy when the invertible multiplier is operated in reverse (as an integer factorizer) over the entire range of factorisable output values (output values which do not have any valid factors which can be represented by two $5$-bit integers are omitted). 
Evident from the test results presented in \cref{table:test_results}, the operating invertible logic circuit converges much faster for prime output values (on average twice as fast, and worst-case execution times are $\approx25\%$ of the general case). In all cases, the input terminals converged to a correct input factor pair for all experiments. 
	\cref{fig:hardware_hist} shows the histograms of convergence cycles in general and prime factorization cases, where the general case includes 340 different outputs ($A \times B = C$) and the prime case includes 66 different outputs.
	The convergence cycles are different depending on the outputs.
	Hence, the worst-case cycle is set to factorize all the numbers.
Fig. \ref{fig:hardware_results} shows measurement results of factorization in arbitrary four selections from all the 340 outputs.

To our knowledge, this work presents the first demonstrations of such invertible logic circuits fabricated on a CMOS process. 
When compared to existing works, where invertible logic circuits were emulated in software on microcontrollers \cite{Pervaiz:pbits}, our results demonstrated latencies several orders of magnitudes faster (tens of $\mu$s, compared to values ranging from hundreds of ms to seconds) than the alternatives.
The reason of the slow speed in \cite{Pervaiz:pbits} is to use software, where the sampling time for p-bits ranges from 1 ms to 400 ms, which would lead to the total latencies of hundreds of ms to seconds.
In contrast, the worst-case latency of the proposed chip are 41.0 $\mu$s.
\begin{table}[t]
    \caption{Invertible Multiplier Circuit Measurement Results}
    \label{table:test_results}
    \centering
    \begin{tabular}{ccc}
        \toprule
        Operation & General factorization & Prime factorization \\
        \midrule
        Manufacturing process & \multicolumn{2}{c}{\acs{TSMC} 65nm \acs{GP}} \\
        \midrule
        Supply voltage & \multicolumn{2}{c}{$1.0\text{V}$} \\
        \midrule
        Clock frequency & \multicolumn{2}{c}{$200 \text{MHz}$} \\
        \midrule
        Power dissipation & \multicolumn{2}{c}{$13.4 \text{mW}$} \\
        \midrule
        Mean cycles & $430$ & $219$ \\
        \midrule
        Mean latency & $2.15 \mu\text{s}$ & $1.10 \mu\text{s}$ \\
        \midrule
        Mean energy & $28.82 \text{nJ}$ & $14.70 \text{nJ}$ \\
        \midrule
        Worst-case cycles & $8192$ & $2048$ \\
        \midrule
        Worst-case latency & $41.0 \mu\text{s}$ & $10.2 \mu\text{s}$ \\
        \bottomrule
    \end{tabular}
\end{table}

\section{Conclusions}
\label{sec:conclusions}

In this work, we have presented a new form of invertible logic based on stochastic computing capable of operating in two modes: 1) a forward mode, in which inputs are presented and a single, correct output is produced, and 2) a reverse mode, in which the output is fixed and the inputs take on values consistent with the output. The underlying processing elements used to construct the invertible logic circuits are streamlined digital spiking neuron models interconnected in Boltzmann machine configurations. 

Furthermore, it was demonstrated that circuits of our logic family are not only invertible but also synthesizable using existing \ac{EDA} tools. When compared to previous works, the developed method constructed high-level circuit constructs using significantly more compact Boltzmann machine configurations. Combined with the streamlined spiking neurons, when targeting a Xilinx Kintex Ultrascale XCKU040-1FBVA \ac{FPGA}, a synthesized 32-bit \ac{RCA} required as little as $27\%$ of the \acp{LUT} and  $11\%$ of the registers when compared to conventional methods. In addition to \ac{FPGA} synthesis results, our stochastic computing based invertible logic was demonstrated to be well suited for \ac{ASIC} fabrication technologies. Test results are also given for a $5$-bit by $5$-bit invertible multiplier/factorizer fabricated using the \ac{TSMC} 65nm \ac{GP} process (an invertible binary multiplier circuit capable of operating in reverse as either a divider circuit or a factorizer circuit).

Overall, we have demonstrated that our design can not only correctly implement basic gates with invertible capability, but can also be extended to construct invertible stochastic adder and multiplier circuits. Our results demonstrate that not only can stochastic computing be used as a theoretical basis for invertible logic, it can also be manufactured with existing methods, taking advantage of the current state-of-the-art technologies and is fully synthesizable. As the first invertible logic circuits to be manufacturable in standard \ac{CMOS} processes illustrates the ability to design and manufacture circuits capable of invertible operations for more complex behaviours with the end goal being invertible regression circuits in support of efficient on-chip machine learning.
%
%
%
%
%

\section*{Acknowledgements}
\label{sec:acknowledgements}
This work was supported in part by MEXT Brainware LSI Project, JSPS KAKENHI Grant Number JP16K12494, JST PRESTO Grant Number JPMJPR18M5, and a PGS-D scholarship from the NSERC. All simulations were supported by VDEC, the University of Tokyo in collaboration with Cadence Inc. and Synopsys, Inc. Furthermore, special thanks are given to Loren Lugosch for his continued support and contributions throughout the project.


\end{document}